\documentclass[letterpaper, 10 pt, conference]{ieeeconf}  

\IEEEoverridecommandlockouts                              

\usepackage[utf8]{inputenc}     

\usepackage[T1]{fontenc}        

\usepackage[english]{babel}              

\usepackage{graphicx}           
\usepackage{amsmath,amssymb,bbm}    
\usepackage{color}
\usepackage[usenames]{xcolor}   
\usepackage{microtype} 

\usepackage{marginnote}         

\usepackage{mathtools}
\usepackage{siunitx}
\usepackage{pbox}
\usepackage{dsfont}
\usepackage{algorithm}

\allowdisplaybreaks

\usepackage{tikz} 
\usetikzlibrary{arrows}				
\usetikzlibrary{shapes.misc}
\usetikzlibrary{patterns}
\usetikzlibrary{decorations.pathreplacing}
\usetikzlibrary{calc, shadings, shadows, snakes, shapes, arrows,trees,patterns}
\usetikzlibrary{chains, fit, automata, positioning}
\usetikzlibrary{math}
\usetikzlibrary{calc}
\usetikzlibrary{decorations.markings}
\usetikzlibrary{fit}
\usepackage{pgfplots}
\usepackage{marvosym}
\usetikzlibrary{calc}

\colorlet{istorange}{orange}
\colorlet{istgreen}{green!50!black}
\colorlet{istblue}{blue} 
\colorlet{istred}{red!90!black}

\newtheorem{theorem}{Theorem}
\newtheorem{lemma}{Lemma}

\newtheorem{assumption}{Assumption}

\newtheorem{remark}{Remark}

\newtheorem{Algorithm}{\bf Algorithm}

\newcommand{\X}{\mathbb{X}}
\newcommand{\U}{\mathbb{U}}
\newcommand{\R}{\mathbb{R}}
\newcommand{\N}{\mathbb{N}}

\makeatletter
\renewcommand*{\@opargbegintheorem}[3]{\trivlist
	\item[\hskip \labelsep{\textit{ #1\ #2}}] \textit{(#3)}: }
\makeatother

\title{\LARGE \bf Self-triggered MPC robust to bounded packet loss via a min-max approach: extended version$^*$}

\author{Stefan Wildhagen, Matthias Pezzutto, Luca Schenato and Frank Allg{\"o}wer
	\thanks{$^*$F. Allg\"ower is thankful that this work was funded by the Deutsche Forschungsgemeinschaft (DFG, German Research Foundation) under Germany’s Excellence Strategy -- EXC 2075 -- 390740016 and under AL 316/13-2 -- 285825138. L. Schenato is thankful that this work was funded by the Italian Ministry of Education, University and Research, through the PRIN project 2017NS9FEY “Realtime Control of 5G Wireless Networks”. S. Wildhagen thanks the Stuttgart Center for Simulation Science (SimTech) for supporting him. S.~Wildhagen and F.~Allg\"ower are with the University of Stuttgart, Institute for Systems Theory and Automatic Control, Germany. {\tt\small \{wildhagen,}{\tt\small allgower\}@ist.uni-}{\tt\small stuttgart.de}. M.~Pezzutto is with the Université Libre de Bruxelles, Service d’Automatique et d’Analyse des Systèmes. {\tt\small matthias.pezzutto@ulb.be}. L.~Schenato is with the University of Padova, Department of Information Engineering, Italy. {\tt\small schenato@dei.unipd.it}} %
}

\begin{document}

\maketitle
\thispagestyle{empty}
\pagestyle{empty}

\begin{abstract}
Networked Control Systems typically come with a limited communication bandwidth and thus require special care when designing the underlying control and triggering law. A method that allows to consider hard constraints on the communication traffic as well as on states and inputs is self-triggered model predictive control (MPC). In this scheme, the optimal length of the sampling interval is determined proactively using predictions of the system behavior. However, previous formulations of self-triggered MPC have neglected the widespread phenomenon of packet loss, such that these approaches might fail in practice. In this paper, we present a novel self-triggered MPC scheme which is robust to bounded packet loss by virtue of a min-max optimization problem. We prove recursive feasibility, constraint satisfaction and convergence to the origin for any possible packet loss realization consistent with the boundedness constraint, and demonstrate the advantages of the proposed scheme in a numerical example.
\end{abstract}

\section{Introduction} \label{sec:intro}

In Networked Control Systems (NCSs), the traditional dedicated communication links are replaced by shared, digital and possibly wireless communication networks. In order to account for the fact that the bandwidth of these networks is often limited, the recent decades have seen a great interest in event- and self-triggered control (ETC, STC) (see \cite{Heemels12} for an overview), which aim at reducing the amount of communication required by the control. Especially self-triggered model predictive control (MPC) \cite{berglind2012self,henriksson2015multiple} has proven as an effective approach to incorporate explicit constraints on the bandwidth as well as on the state and input. Therein, both the transmitted control and the next sampling time instant, i.e., the next time when the MPC is activated and the control input is updated, are determined based on predictions of the future system behavior.

Another unfavorable effect that NCSs typically experience is packet loss, which may be caused by network congestion, packet transmission failures or excessively long propagation delays \cite{zhang2012network}. In the literature, the packet loss process is often modeled in one of the following two ways: As a Bernoulli process \cite{sinopoli2004kalman}, which is especially suitable for describing packet loss of rather stochastic nature, e.g., due to disturbances in the channels or transmission delays. An alternative, deterministic characterization of packet loss is via an upper bound on the number of consecutive packet dropouts \cite{yue2004state}. Although to the best of our knowledge, there exist no works on self-triggered MPC under packet loss, this topic has received some attention in traditional MPC. While most corresponding works take the stochastic perspective (see, e.g., \cite{fischer2013sequence,mishra2018stabilizing}), a few works considered a bounded number $P\in\N$ of consecutive losses, thereby allowing to give deterministic guarantees on convergence: \cite{munoz2008lyapunov,quevedo2011packetized} suggested to include the first $P+1$ predicted control updates in the transmitted packet, while in \cite{ding2011stabilization} the same input is repeatedly transmitted until it is successfully received. However, both strategies might involve serious drawbacks compared to the transmission of a single, fresh control update, especially when $P$ is large: In the former, either one must accept a higher quantization error, or increase the message length, which in turn leads to a higher bandwidth demand. In the latter, the input might be outdated once it arrives, leading to a poor control performance.

For these reasons, we are aiming for an approach that is robust to bounded packet loss despite transmitting a single control update. We propose to do so via min-max MPC \cite{scokaert1998min,raimondo2009min,liu2018robust}, which is a framework that allows to accommodate disturbances on the controlled system by explicitly considering all possible disturbance realizations in the prediction. Thereby, min-max MPC guarantees constraint satisfaction and convergence to a disturbance invariant set. Our main idea is then to regard the packet loss process as a disturbance acting on the NCS. However, the results in \cite{scokaert1998min,raimondo2009min,liu2018robust}, which consider arbitrary bounded-valued disturbances, are not directly transferable to the problem at hand: If one would dissolve the packet loss process into a disturbance taking arbitrary values in $\{0,1\}$, then never having successful transmissions would be among the possible realizations. This, in turn, would mean that the disturbance invariant set spans the entire state space in the case of open-loop unstable plants, such that constraint satisfaction would be contradicted and the convergence result rendered meaningless.

In this paper, we propose a novel self-triggered MPC scheme that is able to accommodate both explicit constraints on the communication bandwidth as well as a bounded number $P$ of consecutive packet losses. In order to achieve this, first, we comprehend the packet loss process as a time-varying disturbance and we assume that the MPC is informed of the success or failure of the latest packet transmission via a reliable acknowledgment channel. Second, using this information, we take into account all possible future disturbance realizations in the prediction via a min-max optimization problem. Thereby, we require that the predicted sampling instants are such that the bandwidth constraint is fulfilled. Third, in order to give theoretical guarantees, we propose to apply a suitable terminal control law for the last $P$ time steps in the prediction. Finally, we prove recursive feasibility, satisfaction of the constraints on the bandwidth, state and input, and convergence to the origin for any possible realization of the packet loss process.

The remainder of the paper is organized as follows. In Section \ref{sec:setup}, we introduce the considered NCS architecture, we present how we capture the bandwidth constraint, and we formalize the assumption on the packet loss process. In Section \ref{sec:min_max}, we formulate the self-triggered MPC algorithm, we present how to design the terminal ingredients, and we establish recursive feasibility, constraint satisfaction and convergence. A numerical example is provided in Section \ref{sec:num_ex}, where we present the advantages of our proposed scheme compared to nominal self-triggered MPC in the presence of packet loss. We summarize the paper in Section \ref{sec:summary}.

Notation: $\N$ denotes the set of natural numbers, $\N_0\coloneqq \N\cup \{0\}$, $\N_{[a,b]}\coloneqq\N_0\cap[a,b]$, and $\N_{\ge a}\coloneqq\N_0\cap[a,\infty)$, $a,b\in\N_0$. $I$ denotes the identity matrix and $0$ the zero matrix of appropriate dimension. $A\succ0$ $(A\succeq 0)$ denotes a symmetric positive {(semi)} definite matrix $A\in\R^{n\times n}$. We denote by $\land$ the logical \emph{and}. For a function $f:\R^n\rightarrow\R^n$, we define the image of a subset $S\subseteq\R^n$ as $f(S)\coloneqq\{f(x):x\in S\}$. The indicator function of a set $S$ is denoted by $\mathds{1}_S(x)$. We denote the cardinality of a countable set $S$ by $|S|$.

\section{Setup} \label{sec:setup}

\begin{figure}
	\begin{tikzpicture}[>= latex]
			\node [thick,draw,rectangle, inner sep=1.5pt,minimum size=1cm,align = center,rounded corners=3pt] (actuator) at (0.2*7,0) {\small{ZOH Actuator}};
			\node[anchor=south east,inner sep=1pt] at (actuator.south east) {\textcolor{istblue}{$w$}};
			
			\node [thick,draw,rectangle, inner sep=1.5pt,minimum size=1cm,align = center,rounded corners=3pt] (plant) at (0.52*7,0) {\small{LTI Plant}};
			\node[anchor=south east,inner sep=1pt] at (plant.south east) {\textcolor{istblue}{$x$}};
			
			\node [thick,draw,rectangle, inner sep=1.5pt,minimum size = 1cm,align = center,rounded corners=3pt] (ctrl) at (0.93*7,-0.2) {\small{\begin{tabular}{c} Self-trig- \\ gered MPC\end{tabular}}};
			\node [thin,dashed,draw,color=istblue,rectangle,minimum height = 1.5cm, minimum width = 3cm,align = left] (smartsensor) at (0.965*7,-0.3) {\hspace{0.0cm}\\[1.1cm] \hspace{-1.1cm}\small{Smart Sensor}};
			\node [thick,draw,rectangle, inner sep=1.5pt,minimum size = 1cm,align = center,rounded corners=3pt] (netw) at (0.5*7,-2) {\small{Token Bucket TS}};
			\node[anchor=south east,inner sep=1pt] at (netw.south east) {\textcolor{istblue}{$\beta$}};
			\node [thin,dashed,draw,cloud,color=istblue,cloud puffs = 16,minimum height = 2.4cm, minimum width = 4.6cm,align = left] (network) at (0.43*7,-2.1) {};
			\node (netwname) at (0.43*7,-2.8) {\color{istblue}\small{Network}};
			
			\node (left) at (0.67*7,0){};
			\node (right) at (0.725*7,0){};
			\node (upper) at (0.72*7,0.25){};
			
			\path (0,0) --++ (1.13*7,-0.2) node(helpne){} --++ (0,-1.8) node(helpse){} --++ (1.13*-7,0) node(helpsw){} --++ (0,2) node(helpnw){};
			
			\draw[thick] (plant.east) -- (left.center) node[pos = 0.5,above] {$x$} -- (upper.center) node[pos = 0.5,above] {$\Delta$};
			\draw[->,thick] (right.center) -- ($(ctrl.west)+(0,0.2)$);
			\draw [->,thick] (ctrl.east) -- (helpne.center)node[pos=1.2,above] {} -- (helpse.center) -- (netw.east);
			\draw (1.08*7+0.2,0.4) node[anchor=north] {$v$};
			\draw (0,0.4) node[anchor=north] {$v$};
			
			\draw[->,thick] (netw.west) -- (helpsw.center) -- (helpnw.center) -- (actuator.west);
			\draw[->,thick] (actuator.east) -- (plant.west) node[pos = 0.5,above] {$u$};
			
			\node [thick,color=istred,label={[label distance=-3pt]270:\color{istred}\small{Dropout}}] (dropout) at (0.25*7,-2) {\Huge\Lightning};
			
			\draw[->,thick,dashed] (dropout.north) --++ (0,0.8) --++ (2.7,0) --++ (0,0.375) -- ($(ctrl.west)-(0,0.2)$) node[pos=0.5,below] {\hspace{-6pt}\small{ACK}};
\end{tikzpicture}
	\caption{Considered NCS with token bucket TS and packet dropouts.}
	\label{fig:NCS_architecture}
\end{figure}
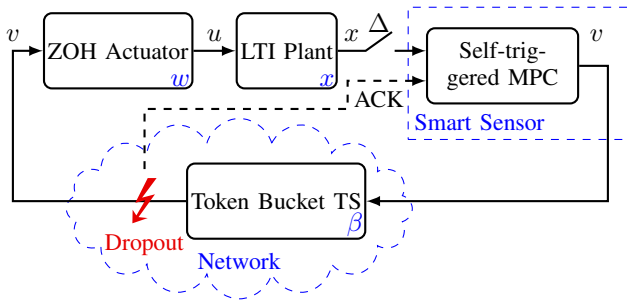

The considered NCS setup is depicted in Figure \ref{fig:NCS_architecture}. A smart sensor measures the state $x$ of the linear time-invariant (LTI) plant and runs the self-triggered MPC. The self-triggered MPC not only decides about the control update $v$, but also about the applied sampling interval $\Delta$, i.e., about the next time when a state measurement is taken, the MPC is activated, and a new control update is computed. The control update $v$ is then transmitted to the actuator via a communication network, and is zero-order held (ZOH) until a new update is received. The network has a limited bandwidth, which is captured by a so-called token bucket traffic specification (TS) \cite{Tanenbaum11}. In addition, the network is prone to packet loss, where we assume that no more than $P\in\N_0$ consecutive packets are lost. Furthermore, we suppose the smart sensor is informed about the success or failure of the latest packet transmission via a reliable acknowledgment channel. In the following, we detail the individual components of the NCS.

\subsection{Plant}

We consider the LTI plant
\begin{equation} \label{eq:plant}
	x(t+1) = A x(t) + B u(t), \; x(0) \in \R^n
\end{equation}
with state $x(t)\in\X\subseteq\R^n$, input $u(t)\in\U\subseteq\R^m$ and where $t\in\N_0$ is the current time. The state and input constraint sets $\X$ and $\U$ are assumed to be closed and contain the origin. For a given $Q,R\succ 0$, the quadratic cost
\begin{equation} \label{eq:cost_plant}
	x^\top Q x + u^\top R u
\end{equation}
is associated with the plant to measure performance.

\subsection{Smart sensor and self-triggered MPC}

At a given sampling instant $t_k, k\in\N_0$, where $t_0=0$, the smart sensor measures the state of the plant $x(t_k)$ and collects the acknowledgement $\text{ACK}(t_k)\in\{0,1\}$, which takes the value $1$ in case of a successful transmission of the packet at $t_{k-1}$, and $0$ in case of packet loss. Then, it activates the self-triggered MPC, which determines both the control update $v(t_k)\in\U$ and the sampling interval $\Delta(t_k)\in\N$ according to the law
\begin{equation} \label{eq:MPC_control_law}
	\begin{bmatrix} v(t_k)^\top & \Delta(t_k)	\end{bmatrix}^\top = \kappa_\text{STMPC}(x(t_k),\text{ACK}(t_k)).
\end{equation}
Thereby, $\kappa_\text{STMPC}:\X\times\{0,1\}\to\U\times\N$ is the self-triggered MPC (STMPC) control law, which is defined implicitly by the solution to an optimization problem detailed in Section \ref{sec:min_max}. Finally, the control update $v(t_k)$ is transmitted over the network and the next sampling instant, at which the entire process is repeated, is determined via $t_{k+1}\coloneqq t_k+\Delta(t_k)$.

\subsection{Network: token bucket TS and bounded packet loss}
\label{sec:network}

As mentioned in Section \ref{sec:setup}, transmission of the control updates $v(t_k)$ must take place over a communication network. We assume that the requested network traffic must fulfill the token bucket TS \cite{Tanenbaum11}, which is based on an analogy to a bucket containing a variable amount of tokens. Tokens are added to the bucket at a constant rate of $g\in\N_{\ge 1}$, while their level is reduced by the cost $c\in\N_{\ge g}$ when a transmission is triggered. The bucket size is $b\in\N_{\ge c}$ and arriving tokens are discarded if the bucket is already full. Hence, the level of tokens in the bucket is governed by
\begin{equation} \label{eq:bucket}
	\beta(t+1)=\begin{cases} \min\{\beta(t)+g-c,b\} & t=t_k \\
							 \min\{\beta(t)+g,b\} & t\in\N_{[t_k+1,t_{k+1}-1]}
			   \end{cases},
\end{equation}
$\beta(0)\in\N_{[0,b]}$. A certain sampling interval can only be chosen by the self-triggered MPC if the tokens are not depleted, i.e., the bucket level is constrained by $\beta(t)\ge 0$ for all $t\in\N_0$. Hence, the average transmission rate allowed by the token bucket TS is $\frac{g}{c}$, and it can be guaranteed a priori that transmissions are possible at a base period of $\lceil\frac{c}{g}\rceil\eqqcolon M$ time instants. Further, the burst length is limited by $\lfloor \frac{b}{c}\rfloor$. Thus, if the parameters $g$, $c$ and $b$ are suitably adjusted, the transmissions will respect both the average and maximum bandwidth of the network. Therefore, in the remainder of the paper, we will presume that the bandwidth constraints are fulfilled if the token bucket TS is fulfilled.

In addition, we consider the possibility that a packet transmission is unsuccessful. We assume that if a packet transmission is successful, it is transmitted instantaneously, and if it is unsuccessful, the packet is lost. Whether the packet containing $v(t_k)$ is lost or not is described by the binary packet loss variable
\begin{equation*}
	\sigma(t_k) = \begin{cases} 1 & \text{the packet is successfully transmitted at } t_k \\
								0 & \text{the packet is lost at } t_k
				  \end{cases}.
\end{equation*}
The received acknowledgment at the smart sensor is then $\text{ACK}(t_k) = \sigma(t_{k-1})$, i.e., the acknowledgment is delivered in at most one sampling interval. Let $\mathcal{T}\coloneqq \{t_{k_1}, t_{k_2}, \ldots\}$ denote the sequence of sampling instants at which the transmission is successful, i.e., at which $\sigma(t_{k_j}) = 1$, $j\in\N_0$.
\begin{assumption} \label{ass:dropout_constraint}
	The number of consecutive packet losses is upper bounded by $P\in\N_0$, i.e., for all $j\in\N_0$, it holds that $k_{j+1}-k_j\in\N_{[1,P+1]}$.
\end{assumption}

\subsection{ZOH Actuator}

The actuator saves the last received control update $v$ in an auxiliary state $w$, which evolves according to
\begin{equation} \label{eq:auxiliary}
	w(t_{k+1}) = \sigma(t_k)v(t_k) + (1-\sigma(t_k))w(t_k).
\end{equation}
Then, depending on whether the packet transmission is successful or not, either the received or the saved control update is applied to the plant
\begin{equation} \label{eq:plant_input}
	u(t) = \sigma(t_k)v(t_k) + (1-\sigma(t_k))w(t_k), \; t\in\N_{[t_k, t_{k+1}-1]}.
\end{equation}

\subsection{Overall NCS}

In the previous subsections, we have observed that, next to the plant state $x$, the saved input $w$ as well as the bucket level $\beta$ represent a dynamical state of the overall NCS. For this reason, we define the overall NCS state
\begin{equation*}
	\xi\coloneqq\begin{bmatrix} x^\top & w^\top & \beta	\end{bmatrix}^\top.
\end{equation*}
Similarly, the overall input consists of the control updates and the chosen sampling interval
\begin{equation*}
	\pi\coloneqq\begin{bmatrix} v^\top & \Delta	\end{bmatrix}^\top.
\end{equation*}
The cost associated with the plant \eqref{eq:cost_plant} can then be written in dependence of the overall state $\xi$, the overall input $\pi$ and the packet loss variable $\sigma$
\begin{equation*}
	\ell(\xi,\pi,\sigma) \coloneqq x^\top Q x + (1-\sigma)w^\top R w + \sigma v^\top R v.
\end{equation*}
The overall state and input constraints are then $\xi(t)\in\Xi\coloneqq\X\times\U\times\N_{[0,b]}$ and $\pi(t)\in\Pi\coloneqq\U\times\N_{[1,\overline{\Delta}]}$. The sampling interval upper bound $\overline{\Delta}\in\N$ is not required for theoretical guarantees, but facilitates evaluating the self-triggered MPC control law $\kappa_\text{STMPC}$ in practice.

\section{Self-triggered Min-Max MPC} \label{sec:min_max}

In order to formulate the self-triggered MPC algorithm, first we describe the dynamics of the overall state, when a control update $v$ is held constant for $j\in\N$ time steps, in a compact form. Denoting $A_j\coloneqq A^j$, $B_j \coloneqq \sum_{i=0}^{j-1} A^i B$, we define
\begin{equation*}
	f(\xi,\begin{bmatrix} v \\ j \end{bmatrix},\sigma) \coloneqq \begin{bmatrix}
		A_j x + (1-\sigma) B_j w + \sigma B_j v \\
		(1-\sigma) w + \sigma v \\
		\min\{\beta + j g - \mathds{1}_{\{\Delta\}}(j)c, b\}
	\end{bmatrix}.
\end{equation*}
Then, according to \eqref{eq:plant}, \eqref{eq:bucket}, \eqref{eq:auxiliary} and \eqref{eq:plant_input}, the overall state $\xi$ at sampling instants evolves according to
\begin{equation*}
	\xi(t_{k+1}) = f(\xi(t_k),\pi(t_k),\sigma(t_k))
\end{equation*}
for all $k\in\N_0$, and in between sampling instants satisfies
\begin{equation*}
	\xi(t_k+j) = f(\xi(t_k),\begin{bmatrix} v(t_k) \\ j \end{bmatrix},\sigma(t_k))
\end{equation*}
for all $j\in\N_{[1,\Delta(t_k)-1]}=\N_{[1,t_{k+1}-t_k-1]}$, for all $k\in\N_0$. We also define the cumulated cost over one sampling interval
\begin{equation*}
	\lambda(\xi,\pi,\sigma) \coloneqq \ell(\xi,\pi,\sigma) + \sum_{j=1}^{\Delta-1} \ell(f(\xi,\begin{bmatrix} v \\ j
	\end{bmatrix},\sigma),\pi,\sigma).
\end{equation*}
Then, it holds for all $k\in\N_0$
\begin{equation*}
	\lambda(\xi(t_k),\pi(t_k),\sigma(t_k)) = \sum_{j=0}^{\Delta(t_k)-1} \ell(\xi(t_k+j),\pi(t_k),\sigma(t_k)).
\end{equation*}

The main idea is now to consider all those packet loss sequences, which are consistent with Assumption \ref{ass:dropout_constraint} and the past realization of the loss sequence, in the prediction. The prediction horizon is given by $N+P$ for an arbitrary $N\in\N$, which will be explained in the following paragraph. Let us first introduce the counter $r(t_k)$, which indicates how long ago was the last sampling instant with a successful transmission. The counter is initialized with $r(t_0)=0$ and evolves according to $r(t_k)=0$ if $\sigma(t_{k-1})=\text{ACK}(t_k)=1$, i.e., the last transmission was successful, and $r(t_k) = r(t_{k-1})+1$ if $\sigma(t_{k-1})=\text{ACK}(t_k)=0$, i.e., the packet was lost. Hence, at sampling instant $t_k$, the MPC is aware that the last successful transmission was at $t_{k-r(t_k)-1}$. Then, we consider in the prediction at $t_k$ all those loss sequences which lie in the set $\mathbb{D}_{N+P}^P(r(t_k))$. This set is defined as the set of all loss sequences of length $N+P$, for which the first successful transmission occurs after at most $P-r(t_k)$ time steps, and the remaining successful transmissions are no further than $P+1$ steps apart. To introduce a formal definition of this set, let us consider a loss sequence $\sigma(\cdot)\in\{0,1\}^{N+P}$ and denote by $\tau_{\sigma(\cdot)}(\cdot)$ the sequence of time indices for which $\sigma(i)=1$, $i\in\N_{[0,N+P-1]}$. For example, if $\sigma(\cdot) = \{0,1,1,0,0,1,\ldots\}$, then $\tau_{\sigma(\cdot)}(\cdot)=\{1,2,5,\ldots\}$. For an $r\in\N_0$, the admissible loss sequence set is then defined by
\begin{equation*}
	\begin{aligned}
		&\mathbb{D}_{N+P}^P(r) = \{\sigma(\cdot)\in\{0,1\}^{N+P} | \tau_{\sigma(\cdot)}(0)\le P-r \\
		&\land \tau_{\sigma(\cdot)}(i+1)-\tau_{\sigma(\cdot)}(i)\le P+1, \; \forall i\in\N_{[0,\lvert\tau_{\sigma(\cdot)}(\cdot)\rvert-2]} \\
		&\land N+P-\tau_{\sigma(\cdot)}(\lvert\tau_{\sigma(\cdot)}(\cdot)\rvert) \le P + 1 \}.
	\end{aligned}
\end{equation*}

Now, we are able to state the optimization problem solved by the self-triggered MPC at each sampling instant. As we have already mentioned in the previous paragraph, the prediction horizon is chosen as $N+P$ with $N\in\N$. This relates to the special feature that in the first part of the horizon from $0$ to $N-1$, the predicted control updates and sampling intervals can be freely chosen, whereas in the second part from $N$ to $N+P-1$, a terminal control law is applied\footnote{This is similar to many classical formulations of MPC, where the prediction and ``control'' horizon are considered separately (see, e.g., \cite{garcia1989model}).}. At a sampling instant $t_k$, we denote by $\sigma(\cdot|t_k)\coloneqq\{\sigma(0|t_k), \ldots,\sigma(N+P-1|t_k) \}\in\mathbb{D}_{N+P}^P(r(t_k))$ an admissible predicted packet loss sequence. Furthermore, consider the sequence of predicted free controls and states
\begin{equation} \label{eq:feedback_policy}
	\begin{aligned}
		&\kappa(\cdot|t_k)=\{\kappa(0|t_k), \ldots,\kappa(N-1|t_k)\} \\
		&\hspace{-1pt}\coloneqq\left\{\hspace{-1pt}\begin{bmatrix} K(0|t_k)x(0|t_k) \\ \Delta(0|t_k) \end{bmatrix}, \ldots, \begin{bmatrix} K(N-1|t_k)x(N-1|t_k) \\ \Delta(N-1|t_k)	\end{bmatrix}\hspace{-1pt}\right\}, \\
		&\xi(\cdot|t_k) \coloneqq \{\xi(0|t_k), \ldots,\xi(N+P|t_k)\}.
\end{aligned}
\end{equation}
As in many approaches in min-max MPC, in the prediction we consider the feedback control policy $\kappa$ instead of an open-loop control, which may lead to a larger region of attraction and enables guaranteed recursive feasibility \cite{raimondo2009min}. As seen in \eqref{eq:feedback_policy}, we demand a linear state feedback for the control updates. We define the objective function by
\begin{equation*}
	\begin{aligned}
		&V(\xi(\cdot|t_k),\kappa(\cdot|t_k),\sigma(\cdot|t_k)) \hspace{-1pt}\coloneqq\hspace{-1pt} \sum_{i=0}^{N-1} \lambda(\xi(i|t_k),\kappa(i|t_k),\sigma(i|t_k)) \\ &+\hspace{-3.5pt}\sum_{i=N}^{N+P-1}\hspace{-3.5pt} \lambda(\xi(i|t_k),\kappa_f(\xi(i|t_k)),\sigma(i|t_k))\hspace{-1pt} +\hspace{-1pt} V_f(\xi(N+P|t_k)),
	\end{aligned}
\end{equation*}
where $\kappa_f:\Xi_f\to\Pi$ is the terminal control law, $\Xi_f\subseteq\Xi$ is the terminal set and $V_f:\Xi_f\to\R$ is the terminal cost. These terminal ingredients will be detailed in Subsection \ref{sec:terminal_ingredients}.

Finally, we define the min-max optimization problem. At each sampling instant $t_k$, given the overall state $\xi(t_k)$ and the counter $r(t_k)$, the self-triggered MPC solves
\begin{subequations} \label{eq:MPC_OP}
	\begin{align}
		&\min_{\xi(\cdot|t_k),\kappa(\cdot|t_k)} \max_{\sigma(\cdot|t_k)} V(\xi(\cdot|t_k),\kappa(\cdot|t_k),\sigma(\cdot|t_k))  \nonumber \\
		\text{s.t. } &\xi(0|t_k)=\xi(t_k), \label{constr_IC} \\
		&\xi(i+1|t_k) = f(\xi(i|t_k),\kappa(i|t_k),\sigma(i|t_k)), \label{constr_dynamics} \\
		&\xi(i|t_k) \in\Xi, \;\kappa(i|t_k) \in \Pi, \; \forall i\in\N_{[0,N-1]}, \label{constr_input}  \\
		&\sigma(\cdot|t_k)\in\mathbb{D}_{N+P}^P(r(t_k)), \label{constr_packet_loss}\\
		&f(\xi(i|t_k),\begin{bmatrix} K(i|t_k)x(i|t_k) \\ j	\end{bmatrix},\sigma(i|t_k))\in\Xi, \nonumber \\ &\forall j\in\N_{[1,\Delta(i|t_k)-1]}, \; \forall i\in\N_{[0,N-1]}, \label{constr_state_inter}\\
		&\xi(i+1|t_k) = f(\xi(i|t_k),\kappa_f(\xi(i|t_k)),\sigma(i|t_k)), \label{constr_dynamics_terminal} \\
		&\xi(i|t_k) \in \Xi_f, \; \forall i\in\N_{[N,N+P]}. \label{constr_terminal}
	\end{align}
\end{subequations}

We denote the min-max optimization problem \eqref{eq:MPC_OP} by $\mathcal{P}(\xi(t_k),r(t_k))$ and by the superscript $^*$ the state, feedback and packet loss sequences that solve $\mathcal{P}$. In addition to the typical constraints on initial condition \eqref{constr_IC}, dynamics \eqref{constr_dynamics},\eqref{constr_dynamics_terminal} and state and input constraint satisfaction \eqref{constr_input}, we require that the predicted packet loss sequences lie in the admissible set \eqref{constr_packet_loss}, the state constraints are satisfied also in between sampling instants \eqref{constr_state_inter}, and the predicted state lies in the terminal set $\Xi_f$ for the last $P+1$ prediction steps \eqref{constr_terminal}. The self-triggered MPC then operates according to Algorithm \ref{scheme_MPC}. Hence, the STMPC control law \eqref{eq:MPC_control_law} is given by
\begin{equation*}
	\begin{aligned}
		\begin{bmatrix} v(t_k)^\top & \Delta(t_k) \end{bmatrix}^\top &= \pi(t_k)= \kappa^*(0|t_k) = \begin{bmatrix} K^*(0|t_k)x(t_k) \\ \Delta^*(0|t_k) \end{bmatrix} \\ &\eqqcolon \kappa_\text{STMPC}(x(t_k),\text{ACK}(t_k)).
	\end{aligned}
\end{equation*}

\begin{algorithm}[h]
	\begin{Algorithm}\label{scheme_MPC}
		\normalfont{\textbf{Self-triggered min-max MPC algorithm}}
		\begin{enumerate}
			\item[0)] Set $t_0=0$, $k=0$ and $r(t_0)=0$.
			\item[1)] If $k\ge 1$, collect $\text{ACK}(t_k)$. If $\text{ACK}(t_k)=1$, set $r(t_k)=0$. If $\text{ACK}(t_k)=0$, set $r(t_k)=r(t_{k-1})+1$.
			\item[2)] Measure $\xi(t_k)$ and solve $\mathcal{P}(\xi(t_k),r(t_k))$.
			\item[3)] Transmit $v(t_k) = K^*(0|t_k)x(t_k)$ over the network.
			\item[4)] Set $t_{k+1} = t_k + \Delta^*(0|t_k)$, $k\leftarrow k+1$ and go to 1).
		\end{enumerate}
	\end{Algorithm}
\end{algorithm}

\begin{remark}
	In the optimization problem \eqref{eq:MPC_OP}, all constraints \eqref{constr_IC}-\eqref{constr_terminal} must be fulfilled for all possible packet loss sequences $\sigma(\cdot|t_k)\in\mathbb{D}_{N+P}^N(r(t_k))$
\end{remark}

\begin{remark}
	In Steps 1) and 2) of Algorithm \ref{scheme_MPC}, the smart sensor only needs to measure the $\text{ACK}$ and $x$, whereas $w$, $\beta$ and $r$ are internal variables. This justifies writing $\kappa_\text{STMPC}$ as a function of the plant state and acknowledgment only.
\end{remark}

\subsection{Design of the terminal ingredients} \label{sec:terminal_ingredients}

In this subsection, we elaborate on how the terminal ingredients should be designed such that theoretical guarantees on recursive feasibility, constraint satisfaction and convergence can be given. We have observed in Subsection \ref{sec:network} that choosing the sampling interval to the base period $\Delta=\lceil\frac{c}{g}\rceil\eqqcolon M$ of the token bucket TS is a feasible strategy. Based on this observation, we consider the terminal control law $\kappa_f(\xi) = \begin{bmatrix} K_f x \\ M \end{bmatrix}$ with a terminal control gain $K_f\in\R^{m\times n}$. For the terminal set, we propose
\begin{equation*} \label{eq:terminal_set}
	\Xi_f\coloneqq \X_f\times\U\times\N_{[c-g,b]},
\end{equation*}
where $\X_f\subseteq\X$ is a set satisfying the following assumption.
\begin{assumption}
	\label{ass:terminal_set}
	The set $\X_f$ is closed and contains the origin. Furthermore, it satisfies $K_f \X_f \subseteq \U$ and
	\begin{align*}
		&(A_{pM}+B_{pM}K_f)\X_f\subseteq \X_f, \;\forall p\in\N_{[1,P+1]}, \\
		&(A_{pM+j}\hspace{-1.5pt}+\hspace{-1.5pt}B_{pM+j}K_f)\X_f\hspace{-1pt}\subseteq\hspace{-1pt}\X, \: \forall j\in\N_{[1,M-1]}, \:\forall p\in\N_{[0,P]}.
	\end{align*}
\end{assumption}
For the terminal cost, we consider
\begin{equation*} \label{eq:terminal_cost}
	V_f(\xi) \coloneqq x^\top P_f x
\end{equation*}
where $P_f=P_f^\top\succ 0\in\R^{n\times n}$.

To give theoretical guarantees in MPC, it is typically assumed that the terminal control law renders the terminal set forward invariant and ensures that the terminal cost decreases. In the following result, we establish that the chosen terminal control law, set and cost indeed satisfy these requirements. To formulate the result, for $i\in\N$ we define $Q_i \coloneqq \sum_{j=0}^{i-1} A_i^\top Q A_i$, $S_i \coloneqq \sum_{j=1}^{i-1} A_i^\top Q B_i$, $R_i \coloneqq iR + \sum_{j=1}^{i-1} B_i^\top Q B_i$, and $\begin{bmatrix} \tilde{Q}_i & \tilde{S}_i \\ \tilde{S}_i^\top & \tilde{R}_i \end{bmatrix} \coloneqq \begin{bmatrix} Q_i & S_i \\ S_i^\top & R_i \end{bmatrix}^{-1}$. Furthermore, we define $f_{1}^{\kappa_f}(\xi) \coloneqq f(\xi,\kappa_f(\xi),1)$ and
\begin{equation*}
		f_{p}^{\kappa_f}(\xi) \coloneqq f(f_{p-1}^{\kappa_f}(\xi),\begin{bmatrix} \cdot \\ M \end{bmatrix},0), \; p\in\N_{[2,P+1]}.
\end{equation*}
The function $f_{p}^{\kappa_f}(\xi)$ represents the overall state after $p$ sampling instants, starting from $\xi$, if the terminal control law is applied and the packet loss sequence is $\{1,0,\ldots,0\}$.

\begin{lemma}
	\label{lem:terminal_ingredients}
Suppose Assumption \ref{ass:terminal_set} holds, $\overline{\Delta}\in\N_{\ge M}$ and there exist matrices $X=X^\top\succ 0\in\R^{n\times n}$ and $Y\in\R^{m\times n}$ such that for all $p\in\N_{[1,P+1]}$
\begin{equation}
	\label{eq:LMI_terminal_cost}
	\begin{bmatrix}
		X & 0 & 0 & A_{pM}+B_{pM}Y \\
		\star & \tilde{Q}_{pM} & \tilde{S}_{pM} & X \\
		\star & \star & \tilde{R}_{pM} & Y \\
		\star & \star & \star & X
	\end{bmatrix} \succeq 0
\end{equation}
is satisfied. Then, with $P_f\coloneqq X^{-1}$ and $K_f\coloneqq YX^{-1}$, it holds that $\kappa_f(\Xi_f)\subseteq\Pi$. Further, we have
\begin{equation} \label{eq:ass_terminal_set_constraint}
	f_{p}^{\kappa_f}(\Xi_f) \subseteq \Xi_f, \; \forall p\in{[1,P+1]},
\end{equation}
and
\begin{equation} \label{eq:ass_terminal_inter_sampling_constraint}
	\begin{aligned}
	&f(\Xi_f,\begin{bmatrix} K_f\X_f \\ j \end{bmatrix},1)\in\Xi, \; \forall j\in\N_{[1,M-1]}, \\
	&f(f_{p}^{\kappa_f}(\Xi_f),\begin{bmatrix} \cdot \\ j \end{bmatrix},0) \subseteq \Xi, \;\forall j\in\N_{[1,M-1]}, \; \forall p\in\N_{[1,P]}. 
	\end{aligned}
\end{equation}
Moreover, for all $p\in\N_{[1,P+1]}$, for all $\xi\in\Xi_f$, it holds that
\begin{equation}
	\label{eq:terminal_cost_decrease}
	\begin{aligned}
		V_f(f_{p}^{\kappa_f}(\xi)) - V_f(\xi) \le &-\lambda(\xi,\kappa_f(\xi),1) \\
		&- \sum_{i=1}^{p-1} \lambda(f_{i}^{\kappa_f}(\xi),\kappa_f(\xi),0).
	\end{aligned}
\end{equation}
\end{lemma}

The proof is provided in Appendix \ref{app:prf_lem}. Lemma \ref{lem:terminal_ingredients} ensures that under the terminal control law, the overall state returns to the terminal set $\Xi_f$ at up to $P+1$ consecutive sampling instants \eqref{eq:ass_terminal_set_constraint}, despite the fact that all transmitted packets except the first are lost. Furthermore, the overall state does not leave the state constraint set $\Xi$ in between the sampling instants \eqref{eq:ass_terminal_inter_sampling_constraint}, and a decrease of the terminal cost $V_f$ is ensured for up to $P+1$ consecutive sampling instants \eqref{eq:terminal_cost_decrease}. Whether \eqref{eq:LMI_terminal_cost}, which is a prerequisite for \eqref{eq:terminal_cost_decrease}, can be satisfied depends on the system matrices $A$ and $B$, the maximum number of packet losses $P$ and the base period $M$ of the token bucket TS. Since \eqref{eq:LMI_terminal_cost} is a linear matrix inequality (LMI), an efficient search for suitable $X$ and $Y$ is possible via a semi-definite program. 

\begin{remark}
		\label{rem:construction_term_regions}
		Suppose the plant state and input constraint sets $\X$ and $\U$ are polytopic. Then, a way to construct $\X_f$ fulfilling Assumption \ref{ass:terminal_set} is the following.
		\begin{enumerate}
			\item Determine an invariant polytope $\X'\subseteq\X$ such that $(A_{pM}+B_{pM}K_f)\X'\subseteq\X'$, for all $p\in\N_{[1,P+1]}$, e.g., via \cite[Algorithm 2]{Pluymers05}; alternatively, define $\X'\coloneqq\{x: x^\top P_f x\le \alpha\}$ for an $\alpha\ge 0$ (note that then, \eqref{eq:QMI_terminal_cost} implies if $x\in \X'$, also $(A_{pM}+B_{pM}K_f)x \in\X'$ for all $p\in\N_{[1,P+1]}$).
			\item Determine $\gamma^* = \max_{\gamma\in[0,\infty)} \gamma$ s.t. $\gamma\X'\subseteq\X$, $\gamma K_f \X'\subseteq\U$ and $\gamma (A_{pM+j}+B_{pM+j}K_f)\X'\subseteq\X$ for all $j\in\N_{[1,M-1]}$, for all $p\in\N_{[0,P]}$.
			\item Define $\X_f\coloneqq\gamma^*\X'$.
		\end{enumerate}
\end{remark}
\begin{remark}
	Note that by virtue of Remark \ref{rem:construction_term_regions}, \eqref{eq:LMI_terminal_cost} is a sufficient condition for existence of an ellipsoidal terminal set satisfying Assumption \ref{ass:terminal_set}.
\end{remark}

\subsection{Theoretical guarantees}

In this subsection, we establish that the proposed self-triggered MPC guarantees recursive feasibility, constraint satisfaction and convergence despite the presence of packet loss. These guarantees depend crucially on the fact that we have required in \eqref{eq:MPC_OP} that the state is contained in the terminal set for the last $P+1$ predicted sampling instants. This ensures that the transmission of at least one of the terminal control laws is successful. This fact, together with Lemma \ref{lem:terminal_ingredients}, warrants that the predicted state can be steered into the terminal region also at $t_{k+1}$ and there is a decrease in the terminal cost. The following result constitutes the main result of this paper and formalizes the guarantees on recursive feasibility, satisfaction of the constraints on bandwith, state and input, and convergence to the origin. The proof is provided in Appendix \ref{app:prf_thm_recursive_feasibility}.

\begin{theorem}
	\label{thm:recursive_feasibility}
	Suppose $\mathcal{P}(\xi(0),0)$ is feasible, Assumptions \ref{ass:dropout_constraint} and \ref{ass:terminal_set} hold, $\overline{\Delta}\in\N_{\ge M}$, there exist matrices $X=X^\top\succ 0\in\R^{n\times n}$ and $Y\in\R^{m\times n}$ such that \eqref{eq:LMI_terminal_cost} holds for all $p\in\N_{[1,P+1]}$, $P_f\coloneqq X^{-1}$, and $K_f\coloneqq YX^{-1}$. Then, $\mathcal{P}$ is feasible for all $t_k$, $k\in\N_0$, and the overall state and input satisfy $\xi(t)\in\Xi$ for all $t\in\N_0$ and $\pi(t_k)\in\Pi$ for all $t_k$, $k\in\N_0$. Moreover, $x(t)$ converges to $0$ as $t\to\infty$.
\end{theorem}
\begin{remark}
	Due to \eqref{eq:plant_input}, Theorem \ref{thm:recursive_feasibility} directly implies that $u(t)\in\U$ for all $t\in\N_0$.
\end{remark}

\section{Numerical Example} \label{sec:num_ex}

In this section, we compare the proposed strategy with a nominal self-triggered MPC, which neglects the possibility of packet loss, and with an oracle self-triggered MPC, which knows the sequence of packet losses in advance. We consider the linearized batch reactor from \cite{Walsh01}, discretized with a sampling period of $\SI{0.1}{\second}$, and a maximum of $P=2$ consecutive packet losses. The parameters of the token bucket TS are $g=1$, $c=3$ and $b=14$, so that $M=3$. We choose the cost matrices $Q=10I$ and $R=I$, a maximum sampling interval of $\overline{\Delta}=5$ and a control horizon of $N=6$. We assume the plant is unconstrained, i.e., $\X=\R^n$ and $\U=\R^m$. We set $\X_f=\R^n$ and obtain $P_f$, $K_f$ according to Lemma~\ref{lem:terminal_ingredients} using YALMIP \cite{YALMIP} and Mosek \cite{MOSEK15}.

In general, min-max MPC schemes are known to be computationally expensive because they require finding the disturbance maximizing the cost along any policy. In the present case, the problem is particularly pronounced because the disturbance takes integer values. In this example, since the plant is unconstrained, we can approach the problem as follows: For any possible sequence of packet losses and sampling periods, we compute offline the optimal sequence of control gains via the finite-horizon LQR problem for time-varying systems. Online, for each pre-computed policy, we evaluate the cost for any packet loss sequence and find the worst case for the given policy. Finally, we choose the policy which gives the minimum worst-case cost.

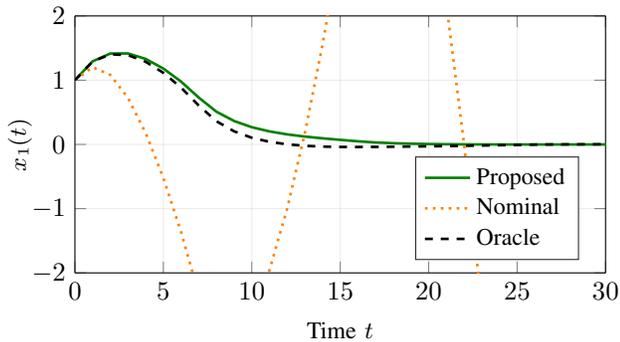
\begin{figure}
%
%
\begin{tikzpicture}

\begin{axis}[%
width=\columnwidth,
height=5cm,
separate axis lines,
xmin=0,
xmax=30,
xlabel style={font=\color{white!10!black},font=\small},
xlabel={Time $t$},
ymin=-2,
ymax=2,
ylabel style={font=\color{white!10!black},font=\small},
ylabel={$x_1(t)$},
ylabel style={yshift=-0.5cm},
axis background/.style={fill=white},
xmajorgrids,
ymajorgrids,
grid style={opacity=0.3},
legend style={at={(0.95,0.05)}, anchor=south east, legend cell align=left, align=left, draw=white!15!black, font=\small},
restrict y to domain=-10:10
]
\addplot [color=istgreen, line width=1.0pt]
  table[row sep=crcr]{%
0	1\\
1	1.29388476854239\\
2	1.41496540402843\\
3	1.41627864692844\\
4	1.33140954617797\\
5	1.18124553610548\\
6	0.978346987674263\\
7	0.729771368888756\\
8	0.509568261939947\\
9	0.364638907250403\\
10	0.268140155171406\\
11	0.202554862929199\\
12	0.15641688592316\\
13	0.122184511122683\\
14	0.0948589799029607\\
15	0.0710864487301798\\
16	0.0485736372939186\\
17	0.0303741275471528\\
18	0.0186078707247344\\
19	0.0109949825035957\\
20	0.00607389128376195\\
21	0.00291124782710289\\
22	0.000914507276463816\\
23	-0.000289053337604766\\
24	-0.000930345456404716\\
25	-0.00114869548163952\\
26	-0.00114806986715923\\
27	-0.00108666486397357\\
28	-0.00098271374486245\\
29	-0.000849948985097783\\
30	-0.00069889406476968\\
31	-0.000537780268146312\\
32	-0.000373204135309813\\
33	-0.000210604330403081\\
34	-5.46119444736851e-05\\
35	6.1252113009696e-05\\
36	0.000122157448866364\\
37	0.000146074958025064\\
38	0.000145272896469362\\
39	0.000128219746069318\\
40	0.000100844550076518\\
41	6.73751035233341e-05\\
42	3.82639623166991e-05\\
43	1.90339724820427e-05\\
44	6.67802548787469e-06\\
45	-7.70716120649665e-07\\
46	-4.59149891454989e-06\\
47	-5.6066020490216e-06\\
48	-5.43367254679531e-06\\
49	-5.21243516468531e-06\\
50	-4.85532336318203e-06\\
};
\addlegendentry{Proposed}

\addplot [color=istorange, dotted, line width=1.0pt]
  table[row sep=crcr]{%
0	1\\
1	1.2029664406118\\
2	1.0822272710138\\
3	0.723181748121331\\
4	0.178232618913101\\
5	-0.521925131481652\\
6	-1.36101714247721\\
7	-2.33250058614331\\
8	-2.98780829117883\\
9	-3.03925046564995\\
10	-2.65967373594358\\
11	-1.95743837867263\\
12	-0.998611490604405\\
13	0.178673104247753\\
14	1.55494157410863\\
15	3.12341235604559\\
16	4.88617917400232\\
17	6.03906949045464\\
18	6.07444625814674\\
19	5.31911654545035\\
20	3.97986802479709\\
21	2.18474232485899\\
22	0.00978639644626078\\
23	-2.50363239170671\\
24	-5.33764776249512\\
25	-8.49071175099301\\
26	-10.5584640561353\\
27	-10.6766724263896\\
28	-9.44123071375738\\
29	-7.23506284732068\\
30	-4.30136754485968\\
31	-0.791222694472071\\
32	3.20548669210584\\
33	7.63999617491993\\
34	12.4916646231719\\
35	15.6858091989949\\
36	16.0099505566917\\
37	14.4152096211077\\
38	11.5289508074988\\
39	7.76613985835836\\
40	3.4020902207448\\
41	-0.0522946319061514\\
42	-1.83630137403353\\
43	-2.6553674271832\\
44	-3.00436518211643\\
45	-3.2401510460589\\
46	-3.62965845774946\\
47	-3.84963545843811\\
48	-3.43106603012587\\
49	-2.32955147256129\\
50	-0.458550546582002\\
};
\addlegendentry{Nominal}

\addplot [color=black, dashed, line width=1.0pt]
table[row sep=crcr]{%
	0	1\\
	1	1.29052035904742\\
	2	1.40249585317884\\
	3	1.38995319164354\\
	4	1.28699236040901\\
	5	1.11469847214735\\
	6	0.885612743248795\\
	7	0.606619903099877\\
	8	0.361645006635615\\
	9	0.204201166315826\\
	10	0.103749614114553\\
	11	0.0404720769875557\\
	12	0.0015087837585073\\
	13	-0.0214820238900391\\
	14	-0.0339074207570519\\
	15	-0.0392671055795492\\
	16	-0.0398199898705638\\
	17	-0.0381153856834255\\
	18	-0.035764253899067\\
	19	-0.0329041209167465\\
	20	-0.0296148801887452\\
	21	-0.0259380876846014\\
	22	-0.0218893708516003\\
	23	-0.0174663619575634\\
	24	-0.012653717844931\\
	25	-0.00742623763627376\\
	26	-0.00308810464369711\\
	27	-0.000443875600776619\\
	28	0.00107894734766994\\
	29	0.00185366348037363\\
	30	0.00212363697137252\\
	31	0.00204734726634572\\
	32	0.00172770530243892\\
	33	0.00123114054876411\\
	34	0.000600033907886541\\
	35	6.3499582032225e-05\\
	36	-0.000238166526208998\\
	37	-0.00038074982236446\\
	38	-0.000413302822208186\\
	39	-0.000367396821107399\\
	40	-0.000263136043184916\\
	41	-0.00015956936371444\\
	42	-9.41565001851098e-05\\
	43	-5.35146066761639e-05\\
	44	-2.90166186533827e-05\\
	45	-1.51138619743845e-05\\
	46	-8.25025932647045e-06\\
	47	-4.92395775044335e-06\\
	48	-2.87207081223923e-06\\
	49	-1.62349992309995e-06\\
	50	-8.78636387295161e-07\\
};
\addlegendentry{Oracle}

\end{axis}
\end{tikzpicture}%
	\vspace*{-5pt}
	\caption{Evolution of the first state.}
	\label{fig:state}
\end{figure}

\begin{figure}
%
%
\begin{tikzpicture}

\begin{axis}[%
width=\columnwidth,
height=4.5cm,
separate axis lines,
xmin=0,
xmax=30,
xlabel style={font=\color{white!10!black},font=\small},
xlabel={Time $t$},
ymin=0,
ymax=6,
ylabel style={font=\color{white!10!black},font=\small},
ylabel={Sampling interval $\Delta(t_k)$},
ylabel style={yshift=-0.5cm},
axis background/.style={fill=white},
xmajorgrids,
ymajorgrids,
grid style={opacity=0.3},
legend style={at={(0.315,0.95)}, anchor=north east, legend cell align=left, align=left, draw=white!15!black, font=\small, legend columns=1},
restrict y to domain=-10:10
]
\addplot[ycomb, color=istgreen, line width=1.0pt, mark=o, mark options={solid, istgreen}] table[row sep=crcr] {%
0	1\\
7	3\\
16	3\\
25	3\\
34	3\\
43	4\\
};
\addlegendentry{Success}
\addplot[ycomb, color=istred, line width=1.0pt, mark=x, mark options={solid, istred}] table[row sep=crcr] {%
1	3\\
4	3\\
10	5\\
15	1\\
19	5\\
24	1\\
28	4\\
32	2\\
37	4\\
41	2\\
47	2\\
49	5\\
};
\addlegendentry{Failure}
\end{axis}
\end{tikzpicture}%
	\vspace*{-5pt}
	\caption{Sampling intervals chosen by proposed self-triggered MPC.}
	\label{fig:sampling_interval}
\end{figure}
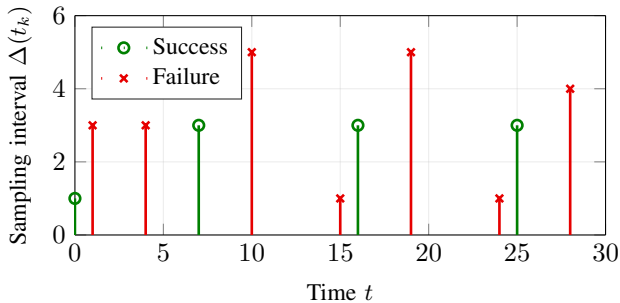

Figure~\ref{fig:state} reports the evolution of the first state. The initial conditions are $x(0)=[1,0,1,0]^\top$, $\beta(0)=8$, and the results were obtained with the packet loss sequence $\sigma(\cdot)=\{1,0,0,1,0,0,1,0,0,\dots\}$. We can see that the nominal self-triggered MPC fails to stabilize the plant under packet loss. In contrast, the proposed self-triggered MPC achieves satisfactory performance despite the high number of lost packets and steers the plant state towards the origin. Interestingly, its performance is only sightly worse than the oracle self-triggered MPC. Figure~\ref{fig:sampling_interval} shows the sampling intervals provided by the proposed strategy, where we can see the varying lengths of the allocated sampling intervals.

\section{Summary and outlook} \label{sec:summary}

In this paper, we have proposed a self-triggered MPC that both allows to incorporate explicit bandwidth constraints and is robust to bounded packet loss. We have achieved this by leveraging the token bucket traffic specification and by considering all possible future loss realizations in the prediction by means of a min-max optimization problem. We have established recursive feasibility, constraint satisfaction and convergence to the origin despite packet loss.

For future work, the presented approach could be adapted to accommodate packet loss also for rollout ETC \cite{Antunes14}. Alternatively, it would be interesting to investigate whether a tube-based approach could achieve the same guarantees. Another promising direction would be to additionally consider a stochastic component of the packet loss process within the proposed framework, such that a wider variety of practical scenarios could be encompassed.

\bibliographystyle{IEEEtran}   
\bibliography{bib_Rollout}

\appendix
\section{Appendix}

\subsection{Proof of Lemma \ref{lem:terminal_ingredients}} \label{app:prf_lem}

Let us begin with the first statement. It is clear that $\kappa_f(\Xi_f)=\begin{bmatrix} K_f\X_f \\ M \end{bmatrix} \stackrel{\text{Ass. \ref{ass:terminal_set}}}{\subseteq} \U\times\N_{[1,\overline{\Delta}]}$. Further, $f_{p}^{\kappa_f}(\Xi_f)=(A_{pM}+B_{pM}K_f)\X_f\times K_f\X_f \times \min\{\N_{[c-g,b]}+Mg-c,b\}\stackrel{\text{Ass. \ref{ass:terminal_set}}}{\subseteq}\X_f\times\U\times \N_{[c-g,b]} =\Xi_f$, which proves \eqref{eq:ass_terminal_set_constraint}. The statement \eqref{eq:ass_terminal_inter_sampling_constraint} follows similarly. To prove \eqref{eq:terminal_cost_decrease}, we apply the Schur complement to \eqref{eq:LMI_terminal_cost} w.r.t. its fourth diagonal element and substitute $X=P_f^{-1}$ and $Y = K_f P_f^{-1}$. By pre- and postmultiplying $P_f$, we obtain
\begingroup
\arraycolsep2pt
\begin{equation}
	\label{eq:QMI_terminal_cost}
	(A_{pM}+B_{pM}K_f)^\top P_f (\star) - P_f + \begin{bmatrix} I \\ K_f \end{bmatrix}^\top\hspace{-2pt} \begin{bmatrix} Q_{pM} & S_{pM} \\ \star & R_{pM}	\end{bmatrix} \star \preceq 0.
\end{equation}
\endgroup
After substituting $Q_{pM}$, $S_{pM}$ and $R_{pM}$ with their definitions, it is obvious that \eqref{eq:QMI_terminal_cost} implies \eqref{eq:terminal_cost_decrease}.

\subsection{Proof of Theorem \ref{thm:recursive_feasibility}} \label{app:prf_thm_recursive_feasibility}

First, we prove recursive feasibility. Suppose at a sampling instant $t_k$, $k\in\N_0$, $\mathcal{P}(\xi(t_k),r(t_k))$ was feasible. For an $r\in\N_0$, denote by $\mathcal{L}_{N+P}^P(r)$ the index set of $\mathbb{D}_{N+P}^P(r)$ and by $\sigma^\ell(\cdot|t_k)$ and $\xi^\ell(\cdot|t_k)$, $\ell\in\mathcal{L}_{N+P}^P(r(t_k))$, the possible packet loss and the corresponding optimal state sequence at $t_k$. Under Algorithm \ref{scheme_MPC}, at time $t_k$, the control law is given by the first element of the optimal feedback $\pi(t_k)=\kappa^*(0|t_k)$, and the packet loss variable takes a certain value $\sigma(t_k)$.

For $r\in\N_0$, $r^+\in\{0,r+1\}$, let us define $\mathbb{D}_{N+P}^P(r,r^+)\coloneqq \mathbb{D}_{N+P}^P(r) \cap \{\sigma(\cdot)\in\{0,1\}^{N+P}|\sigma(0)=1 \text{ if } r^+=0, \; \sigma(0)=0 \text{ if } r^+=r+1\}$ and denote by $\mathcal{L}_{N+P}^P(r,r^+)$ its index set. Hence, $\mathbb{D}_{N+P}^P(r(t_k),r(t_{k+1}))$ is the set of all predicted packet loss sequences which satisfy $\sigma(0|t_k)=\sigma(t_k)$. For this reason, it is obvious that $\xi(t_{k+1})=\xi^\ell(1|t_k)$ for all $\ell\in\mathcal{L}_{N+P}^P(r(t_k),r(t_{k+1}))$. Furthermore, the set $\mathbb{D}_{N+P}^P(r(t_{k+1}))$ contains all those packet loss sequences that are compatible with the past loss realization. For this reason, we have the fact that for each dropout sequence $\sigma^m(\cdot|t_{k+1})\in\mathbb{D}_{N+P}^P(r(t_{k+1}))$, i.e., for all $m\in\mathcal{L}_{N+P}^P(r(t_{k+1}))$, there exists $\ell\in\mathcal{L}_{N+P}^P(r(t_k),r(t_{k+1}))$ such that $\sigma^m(i|t_{k+1})=\sigma^\ell(i+1|t_k)$ for all $i\in\N_{[0,N+P-2]}$.

Now, consider the candidate feedback policy at $t_{k+1}$
\begin{equation*}
	\begin{aligned}
		&\tilde{\kappa}(\cdot|t_{k+1})=\{\tilde{\kappa}(0|t_{k+1}),\ldots,\tilde{\kappa}(N-1|t_{k+1})\} \\
		&\coloneqq\bigg\{\kappa^*(1|t_k), \ldots, \kappa^*(N-1|t_k), \begin{bmatrix} K_f \tilde{x}^m(N-1|t_{k+1}) \\ M	\end{bmatrix}\bigg\}
	\end{aligned}
\end{equation*}
and the corresponding state sequences $\tilde{\xi}^m(\cdot|t_{k+1})$, $m\in\mathcal{L}_{N+P}^P(r(t_{k+1}))$. Note that $\tilde{\kappa}$ lies in the class of feedback policies defined by \eqref{eq:feedback_policy}. With this candidate policy, the fact mentioned in the paragraph above and since $\xi(t_{k+1})=\xi^\ell(1|t_k)$ for all $\ell\in\mathcal{L}_{N+P}^P(r(t_k),r(t_{k+1}))$, it holds that for all $m\in\mathcal{L}_{N+P}^P(r(t_{k+1}))$, there exists $\ell\in\mathcal{L}_{N+P}^P(r(t_k),r(t_{k+1}))$ such that $\tilde{\xi}^m(i|t_{k+1})=\xi^\ell(i+1|t_k)$ for all $i\in\N_{[0,N+P-1]}$. This immediately implies that $\tilde{\xi}^m(0|t_{k+1})$ satisfies \eqref{constr_IC} for all $m$.

For the same reasons, it is immediate that $\tilde{\xi}^m(i|t_{k+1})=\xi^\ell(i+1|t_k)\stackrel{\eqref{constr_input}}{\in}\Xi$, for all $i\in\N_{[0,N-1]}$. As $x^\ell(N|t_k)\in\X_f$ for all $\ell\in\mathcal{L}_{N+P}^P(r(t_k),r(t_{k+1}))$, it also holds that $K_f \tilde{x}^m(N-1|t_{k+1})=K_fx^\ell(N|t_k)\in K_f\X_f\subseteq\U$ from Assumption \ref{ass:terminal_set}. This implies together with $M\le\overline{\Delta}$ that $\tilde{\kappa}(N-1|t_{k+1})\in\Pi$. Obviously, it also holds that $\tilde{\kappa}(i|t_{k+1})=\kappa^*(i+1|t_k)\in\Pi$ for all $i\in\N_{[0,N-2]}$. These facts together imply that $\tilde{\xi}^m(\cdot|t_{k+1})$ and $\tilde{\kappa}(\cdot|t_{k+1})$ satisfy \eqref{constr_input} for all $m$.

Again for the same reasons, it is apparent that $\tilde{\xi}^m(\cdot|t_{k+1})$ satisfies the inter-sampling state constraint \eqref{constr_state_inter} for all $i\in\N_{[0,N-2]}$. As $\tilde{\kappa}(N-1|t_{k+1})$ is the terminal control law and due to \eqref{eq:ass_terminal_inter_sampling_constraint} in Lemma \ref{lem:terminal_ingredients}, \eqref{constr_state_inter} is satisfied also at $i=N-1$.

Next, we turn to the terminal state constraint \eqref{constr_terminal}. We have that for all $m\in\mathcal{L}_{N+P}^P(r(t_{k+1}))$, there exists $\ell\in\mathcal{L}_{N+P}^P(r(t_k),r(t_{k+1}))$ such that $\tilde{\xi}^m(i|t_{k+1})=\xi^\ell(i+1|t_k)\stackrel{\eqref{constr_terminal}}{\in}\Xi_f$ for all $i\in\N_{[N,N+P-1]}$. For $i=N+P$, we distinguish two cases: \\
\underline{Case 1): $\sigma^m(N+P-1|t_{k+1})=1$.} Since $\tilde{\xi}^m(N+P-1|t_{k+1})=\xi^\ell(N+P|t_k)\in\Xi_f$, the terminal controller $\kappa_f(\tilde{\xi}^m(N+P-1|t_{k+1}))$ is applied, and in this case, the corresponding packet is not lost, it clearly holds that $\tilde{\xi}^m(N+P|t_{k+1})=f_1^{\kappa_f}(\xi^\ell(N+P|t_k))$. Thus, with \eqref{eq:ass_terminal_set_constraint} in Lemma \ref{lem:terminal_ingredients}, we have $\tilde{\xi}^m(N+P|t_{k+1})\in\Xi_f$. \\
\underline{Case 2): $\sigma^m(N+P-1|t_{k+1})=0$.} In this case, there must exist an $i_m'\in\N_{[N-1,N+P-2]}$ such that $\sigma^m(i_m'|t_{k+1})=1$, which is simply from the fact that $\sigma^m(\cdot|t_{k+1})\in\mathbb{D}_{N+P}^P(r(t_{k+1}))$. Denote by $i_m''\coloneqq\max i_m'$ the time index of the last successful transmission. Note that in this case, the last successful transmission in $\sigma^\ell(\cdot|t_k)$ was at $i_\ell''=i_m''+1$. As $\tilde{\xi}^m(i_m''|t_{k+1})=\xi^\ell(i_m''+1|t_k)\in\Xi_f$, $\sigma^m(i_m''|t_{k+1})=1$, $\kappa_f(\tilde{\xi}^m(i_m''|t_{k+1}))$ is applied and $N+P-i_m''\le P+1$, it holds that $\tilde{\xi}^m(N+P|t_{k+1})=f_{N+P-i_m''}^{\kappa_f}(\xi^\ell(i_m''+1|t_k))$. Thus, with \eqref{eq:ass_terminal_set_constraint} in Lemma \ref{lem:terminal_ingredients}, we have $\tilde{\xi}^m(N+P|t_{k+1})\in\Xi_f$. \\
In conclusion, all constraints are fulfilled at $t_{k+1}$ such that recursive feasibility is guaranteed.

Second, constraint satisfaction directly follows from the fact that the control law is $\pi(t_k)=\kappa^*(0|t_k)\stackrel{\eqref{constr_input}}{\in}\Pi$ according to Algorithm \ref{scheme_MPC}, which implies that $\xi(t_k+j) = f(\xi^\ell(0|t_k),\begin{bmatrix} K^*(0|t_k)x^\ell(0|t_k) \\ j \end{bmatrix}), \sigma^\ell(0|t_k))\stackrel{\eqref{constr_input},\eqref{constr_state_inter}}{\in}\Xi$ for all $j\in\N_{[1,\Delta^*(0|t_k)]}$ for all $\ell\in\mathcal{L}_{N+P}^P(r(t_k),r(t_{k+1}))$.

Third, we prove convergence. It holds by optimality and an expansion of the optimal cost at $t_{k+1}$
\begin{equation} \label{eq:prf_conv_lyap_1}
	\begin{aligned}
	&V(\xi^*(\cdot|t_{k+1}),\kappa^*(\cdot|t_{k+1}),\sigma^*(\cdot|t_{k+1})) \\
	&\le \max_{m\in\mathcal{L}_{N+P}^P(r(t_{k+1}))} V(\tilde{\xi}^m(\cdot|t_{k+1}),\tilde{\kappa}(\cdot|t_{k+1}),\sigma^m(\cdot|t_{k+1})) \\
	&= \max_{m\in\mathcal{L}_{N+P}^P(r(t_{k+1}))} \bigg[ V_f(\tilde{\xi}^m(N+P|t_{k+1})) \\
	&+\sum_{i=0}^{N-1}\lambda(\tilde{\xi}^m(i|t_{k+1}),\tilde{\kappa}(i|t_{k+1}),\sigma^m(i|t_{k+1})) \\
	&+\sum_{i=N}^{N+P-1}\lambda(\tilde{\xi}^m(i|t_{k+1}),\kappa_f(\tilde{\xi}^m(i|t_{k+1})),\sigma^m(i|t_{k+1})) \bigg].
\end{aligned}
\end{equation}
In the following, $\ell\in\mathcal{L}_{N+P}^P(r(t_k),r(t_{k+1}))$ is such that $\sigma^m(i|t_{k+1})=\sigma^\ell(i+1|t_k)$ for all $i\in\N_{[0,N+P-2]}$ and $\tilde{\xi}^m(i|t_{k+1})=\xi^\ell(i+1|t_k)$ for all $i\in\N_{[0,N+P-1]}$. Note that $\xi(t_k)=\xi^\ell(0|t_k)$ and $\sigma(t_k)=\sigma^\ell(0|t_k)$ for all $\ell\in\mathcal{L}_{N+P}^P(r(t_k),r(t_{k+1}))$. Thus, we may carefully add $0=\lambda(\xi^\ell(0|t_k),\kappa^*(0|t_k),\sigma^\ell(0|t_k))-\lambda(\xi(t_k),\kappa^*(0|t_k),\sigma(t_k))$ to \eqref{eq:prf_conv_lyap_1} and obtain
\begin{equation} \label{eq:prf_conv_lyap}
	\begin{aligned}
	&V(\xi^*(\cdot|t_{k+1}),\kappa^*(\cdot|t_{k+1}),\sigma^*(\cdot|t_{k+1})) \\
	&\le \max_{m\in\mathcal{L}_{N+P}^P(r(t_{k+1}))} \bigg[\sum_{i=0}^{N-1}\lambda(\xi^\ell(i|t_k),\kappa^*(i|t_k),\sigma^\ell(i|t_k)) \\
	&+\sum_{i=N}^{N+P-1}\lambda(\xi^\ell(i|t_k),\kappa_f(\xi^\ell(i|t_k)),\sigma^\ell(i|t_k)) \\
	&+\lambda(\xi^\ell(N\hspace{-1.5pt}+\hspace{-1.5pt}P|t_k),\kappa_f(\xi^\ell(N\hspace{-1.5pt}+\hspace{-1.5pt}P|t_k)),\sigma^m(N\hspace{-1.5pt}+\hspace{-1.5pt}P\hspace{-1.5pt}-\hspace{-1.5pt}1|t_{k+1})) \\
	&+ V_f(\tilde{\xi}^m(N+P|t_{k+1})) \bigg] -\lambda(\xi(t_k),\kappa^*(0|t_k),\sigma(t_k)).
	\end{aligned}
\end{equation}

For the terminal cost, again, we distinguish two cases:
\underline{Case 1): $\sigma^m(N+P-1|t_{k+1})=1$.} As $\tilde{\xi}^m(N+P|t_{k+1})=f_1^{\kappa_f}(\xi^\ell(N+P|t_k))$, Lemma \ref{lem:terminal_ingredients} directly implies
\begin{align*}
	V_f(\tilde{\xi}^m(N+P&|t_{k+1}))-V_f(\xi^\ell(N+P|t_k)) \\
	&\le-\lambda(\xi^\ell(N+P|t_k),\kappa_f(\xi^\ell(N+P|t_k)),1).
\end{align*}
\underline{Case 2): $\sigma^m(N+P-1|t_{k+1})=0$.} It holds that $\xi^\ell(N+P|t_k)=f_{N+P-i_\ell''}^{\kappa_f} (\xi^\ell(i_\ell''|t_k))$. Thus, \eqref{eq:terminal_cost_decrease} in Lemma \ref{lem:terminal_ingredients} implies
\begin{equation}
	\label{eq:prf_conv_1}
	\begin{aligned}
		&V_f(\xi^\ell(N+P|t_k))-V_f(\xi^\ell(i_\ell''|t_k)) \\
		&\le-\lambda(\xi^\ell(i_\ell''|t_k),\kappa_f(\xi^\ell(i_\ell''|t_k)),1)\\
		&-\sum_{i=1}^{N+P-i_\ell''-1} \lambda(f_{i}^{\kappa_f}\hspace{-1pt}(\xi^\ell(i_\ell''|t_k)),\kappa_f(\xi^\ell(i_\ell''|t_k)),0).
	\end{aligned}
\end{equation}
Further, from $i_m''=i_\ell''-1$ we have $\tilde{\xi}^m(N+P|t_{k+1})=f_{N+P-i_\ell''+1}^{\kappa_f}(\xi^\ell(i_\ell''|t_k))$, such that \eqref{eq:terminal_cost_decrease} in Lemma \ref{lem:terminal_ingredients} implies
\begin{equation}
	\label{eq:prf_conv_2}
	\begin{aligned}
		&V_f(\tilde{\xi}^m(N+P|t_{k+1}))-V_f(\xi^\ell(i_\ell''|t_k)) \\
		&\le-\lambda(\xi^\ell(i_\ell''|t_k),\kappa_f(\xi^\ell(i_\ell''|t_k)),1)\\
		&-\sum_{i=1}^{N+P-i_\ell''} \lambda(f_{i}^{\kappa_f}(\xi^\ell(i_\ell''|t_k)),\kappa_f(\xi^\ell(i_\ell''|t_k)),0).
	\end{aligned}
\end{equation}
Subtracting \eqref{eq:prf_conv_1} from \eqref{eq:prf_conv_2} yields
\begin{equation*}
	\begin{aligned}
		&V_f(\tilde{\xi}^m(N+P|t_{k+1}))-V_f(\xi^\ell(N+P|t_k)) \\
		&\le-\lambda(\xi^\ell(N+P|t_k),\kappa_f(\xi^\ell(i_\ell''|t_k)),0)\\
		&=-\lambda(\xi^\ell(N+P|t_k),\kappa_f(\xi^\ell(N+P|t_k)),0).
	\end{aligned}
\end{equation*}

In summary, we have in both cases
\begin{equation*}
	\begin{aligned}
		&V_f(\xi^\ell(N+P|t_k))\ge V_f(\tilde{\xi}^m(N+P|t_{k+1})) \\
		&+\hspace{-1.5pt}\lambda(\xi^\ell(N\hspace{-1.5pt}+\hspace{-1.5pt}P|t_k),\kappa_f(\xi^\ell(N\hspace{-1.5pt}+\hspace{-1.5pt}P|t_k)),\sigma^m(N\hspace{-1.5pt}+\hspace{-1.5pt}P\hspace{-1.5pt}-\hspace{-1.5pt}1|t_{k+1})).
	\end{aligned}
\end{equation*}
We substitute this estimate in \eqref{eq:prf_conv_lyap} and obtain
\begin{align*}
	&V(\xi^*(\cdot|t_{k+1}),\kappa^*(\cdot|t_{k+1}),\sigma^*(\cdot|t_{k+1}))  \\
	&\le \max_{m\in\mathcal{L}_{N+P}^P(r(t_{k+1}))} \bigg[\sum_{i=0}^{N-1}\lambda(\xi^\ell(i|t_k),\kappa^*(i|t_k),\sigma^\ell(i|t_k))  \\
	&+\sum_{i=N}^{N+P-1}\lambda(\xi^\ell(i|t_k),\kappa_f(\xi^\ell(i|t_k)),\sigma^\ell(i|t_k))  \\
	&+ V_f(\xi^\ell(N+P|t_k)) \bigg] -\lambda(\xi(t_k),\kappa^*(0|t_k),\sigma(t_k)) \\
	&= \max_{\ell\in\mathcal{L}_{N+P}^P(r(t_k),r(t_{k+1}))} \bigg[ \ldots \bigg] -\lambda(\xi(t_k),\kappa^*(0|t_k),\sigma(t_k)) \\
	& \le V(\xi^*(\cdot|t_k),\kappa^*(\cdot|t_k),\sigma^*(\cdot|t_k)) -\lambda(\xi(t_k),\kappa^*(0|t_k),\sigma(t_k)) \\
	&\le V(\xi^*(\cdot|t_k),\kappa^*(\cdot|t_k),\sigma^*(\cdot|t_k)) - x(t_k)^\top Q x(t_k).
\end{align*}
Hence, the optimal cost $V$ decreases unless $x(t_k)=0$. Since $V$ is lower bounded, it must converge to a constant value and hence, $x(t_k)\to 0$ as $k\to\infty$ due to $Q\succ 0$. Convergence of $x(t)\to 0$ as $t\to\infty$ follows directly with \cite[Theorem 2]{nesic1999formulas}.

\end{document}